\documentclass[fleqn,usenatbib]{mnras}

\usepackage{newtxtext,newtxmath}

\usepackage[T1]{fontenc}
\usepackage{ae,aecompl}

\usepackage{graphicx}	
\usepackage{amsmath}	
\usepackage{multicol} 
\usepackage{multirow}
\usepackage{comment}
\usepackage{array}
\usepackage{soul}
\hypersetup{colorlinks=true,citecolor=blue,linkcolor=purple,filecolor=black,runcolor=black,breaklinks=true}
\usepackage{etoolbox}
\usepackage{ae,aecompl}
\usepackage[usenames]{color}
\usepackage{hyperref}
\usepackage{threeparttable}
\usepackage{mathrsfs}
\usepackage{scalerel}
\usepackage{longtable,booktabs,threeparttablex}

\definecolor{purple}{RGB}{160,32,240}

\definecolor{red}{RGB}{225,50,50}

\newcommand{\Muv}{\ensuremath{\mathrm{M}_{\mathrm{UV}}^{ }}}

\newcommand{\Lya}{\ensuremath{\mathrm{Ly}\alpha}}
\newcommand{\zphot}{\ensuremath{z_{\mathrm{phot}}}}

\newcommand{\OIIIHb}{[OIII]+H\ensuremath{\beta}}

\newcommand{\zCII}{\ensuremath{z_{_{\mathrm{[CII]}}}}}

\newcolumntype{P}[1]{>{\centering\arraybackslash}p{#1}}
\newcolumntype{L}[1]{>{\raggedright\arraybackslash}p{#1}}
\newcommand\Tstrut{\rule{0pt}{2.6ex}}         

\title[A Highly-obscured Radio-loud AGN at $z=6.853$]{ALMA Confirmation of an Obscured Hyperluminous Radio-Loud AGN at $z=6.853$ Associated with a Dusty Starburst in the 1.5 deg$^2$ COSMOS Field}

\author[R. Endsley et al.]{
Ryan Endsley$^{1}$\thanks{E-mail: rendsley@email.arizona.edu}, 
Daniel P. Stark$^{1}$, 
Jianwei Lyu$^{1}$,
Feige Wang$^{1}$,
Jinyi Yang$^{1,2}$,
Xiaohui Fan$^{1}$, 
Renske Smit$^{3}$,
\newauthor{
Rychard Bouwens$^{4}$,
Kevin Hainline$^{1}$,
Sander Schouws$^{4}$}
\\
$^{1}$Steward Observatory, University of Arizona, 933 N Cherry Ave, Tucson, AZ 85721 USA\\
$^{2}$Strittmatter Fellow\\
$^{3}$Astrophysics Research Institute, Liverpool John Moores University, 146 Brownlow Hill, Liverpool L3 5RF, United Kingdom\\
$^{4}$Leiden Observatory, Leiden University, NL-2300 RA Leiden, Netherlands\\
}

\date{Accepted XXX. Received YYY; in original form ZZZ}

\pubyear{2022}

\begin{document}
\label{firstpage}
\pagerange{\pageref{firstpage}--\pageref{lastpage}}
\maketitle


\begin{abstract}
We present band 6 ALMA observations of a heavily-obscured radio-loud ($L_{1.4\ \mathrm{GHz}}=10^{25.4}$ W Hz$^{-1}$) AGN candidate at $z_\mathrm{phot}=6.83\pm0.06$ found in the 1.5 deg$^2$ COSMOS field. The ALMA data reveal detections of exceptionally strong [CII]158$\mu$m ($z_\mathrm{[CII]}=6.8532$) and underlying dust continuum emission from this object (COS-87259), where the [CII] line luminosity, line width, and 158$\mu$m continuum luminosity are comparable to that seen from $z\sim7$ sub-mm galaxies and quasar hosts. The 158$\mu$m continuum detection suggests a total infrared luminosity of $9\times10^{12}$ $L_\odot$ with corresponding very large obscured star formation rate (1300 $M_\odot$/yr) and dust mass ($2\times10^9$ $M_\odot$). The strong break seen between the VIRCam and IRAC photometry perhaps suggests that COS-87259 is an extremely massive reionization-era galaxy with $M_\ast\approx1.7\times10^{11}$ $M_\odot$. Moreover, the MIPS, PACS, and SPIRE detections imply that this object harbors an AGN that is heavily obscured ($\tau_{_{\mathrm{9.7\mu m}}}=2.3$) with a bolometric luminosity of approximately $5\times10^{13}$ $L_\odot$. Such a very high AGN luminosity suggests this object is powered by an $\approx$1.6 $\times$ 10$^9$ $M_\odot$ black hole if accreting near the Eddington limit, and is effectively a highly-obscured version of an extremely UV-luminous ($M_{1450}\approx-27.3$) $z\sim7$ quasar. Notably, these $z\sim7$ quasars are an exceedingly rare population ($\sim$0.001 deg$^{-2}$) while COS-87259 was identified over a relatively small field. Future very wide-area surveys with, e.g., \textit{Roman} and \textit{Euclid} have the potential to identify many more extremely red yet UV-bright $z\gtrsim7$ objects similar to COS-87259, providing richer insight into the occurrence of intense obscured star formation and supermassive black hole growth among this population.
\end{abstract}



\defcitealias{Endsley2022_radioAGN}{E22a}

\section{Introduction} \label{sec:intro}

For much of the past two decades, deep imaging surveys with the Hubble Space Telescope (\textit{HST}) have served as key windows on galaxies present in the first billion years of cosmic history \citep[e.g.][]{Giavalisco2004,Bouwens2011_HUDF,Grogin2011,Koekemoer2011,Trenti2011,Ellis2013,Illingworth2013,Lotz2017}. 
To date, these \textit{HST} surveys have enabled the identification of over 1000 Lyman-break galaxies at $z\gtrsim6$ (e.g. \citealt{McLure2013,Finkelstein2015_LF,Oesch2018_z10LF,Ishigaki2018,Bouwens2021_LF}), the bulk of which are faint ($\Muv{} > -20$) star-forming systems thought largely responsible for cosmic reionization \citep{Bouwens2015_reionization,Finkelstein2019,Robertson2022}.
Overlapping deep rest-optical imaging with \textit{Spitzer}/IRAC has shown that these galaxies tend to possess low stellar masses ($M_\ast{} < 10^9\ M_\odot$) and strong rest-optical nebular line emission consistent with young ages \citep[e.g.][]{Labbe2013,Stark2013_NebEmission,Smit2014,Smit2015,Song2016,deBarros2019,Endsley2021_OIII,Stefanon2021_SMF}.
However, while these \textit{HST} surveys can push to extremely deep depths ($J \sim 29$), they tend to cover relatively small areas ($\lesssim$0.01 deg$^2$) which do not statistically sample $z\gtrsim6$ galaxies that are sufficiently bright for detailed characterization with existing facilities ($J \lesssim 25$).
Consequently, our understanding of very early galaxies has long been primarily limited to their rest-UV$+$optical photometric properties.

Within the past few years, several deep wide-area ($>$deg$^2$) optical$+$near-infrared imaging surveys have reached (near) completion \citep{Lawrence2007,McCracken2012,Jarvis2013,Inoue2020,Aihara2022}, providing large samples of UV-bright ($-23 \lesssim \Muv{} \lesssim -21$) Lyman-break $z\gtrsim7$ galaxies \citep{Bowler2017,Bowler2020,Stefanon2019,Endsley2021_OIII,Bouwens2022_REBELS,Harikane2022_LF,Schouws2022_CII}.
This has in turn led to substantial improvements in our understanding of the UV-luminous reionization-era galaxy population. 
Rest-UV spectroscopic follow-up has revealed significant Lyman-alpha emission from several of these systems \citep{Ono2012,Furusawa2016,Endsley2021_LyA,Endsley2022_REBELS,Valentino2022}, while (in some cases) simultaneously uncovering strong large-scale overdensities likely tracing very large ionized bubbles \citep{Endsley2022_bubble}.
ALMA follow-up has also been highly successful in detecting far-IR cooling lines ([CII]158$\mu$m and [OIII]88$\mu$m) from several UV-bright $z\gtrsim7$ galaxies \citep{Smit2018,Hashimoto2019,Bouwens2022_REBELS,Bowler2022,Schouws2022_CII}, with a substantial fraction additionally exhibiting strong far-IR continuum emission indicating substantial obscured star formation activity \citep{Hashimoto2019,Inami2022,Schouws2022_dust}. 

Alongside efforts to study Lyman-break $z\gtrsim7$ galaxies, there has also been a dedicated push to identify and subsequently characterize UV-luminous quasars as well as IR-luminous sub-millimeter galaxies in this epoch.
The 2500 deg$^2$ South Pole Telescope Survey \citep{Everett2020} has resulted in the discovery of a sub-mm system at $z=6.900$ (SPT-0311; \citealt{Strandet2017}), the only such source currently known at $z>6.5$. 
ALMA follow-up revealed that this object is actually a merger of two galaxies exhibiting intense dust-obscured star formation activity ($\approx$3500 $M_\odot$/yr combined) that likely traces one of the most massive halos in the reionization era ($\sim$3$\times$10$^{12}$ $M_\odot$; \citealt{Marrone2018,Spilker2022}).
Thanks to a rich collection of $\gtrsim$1000 deg$^2$ optical$+$near-IR surveys, there are now more than 40 known extremely UV-luminous ($-28 \leq \Muv{} \leq -25$) quasars at $z>6.5$ (\citealt{Mortlock2011,Venemans2013,Venemans2015,Mazzucchelli2017,Reed2017,Wang2017,Wang2018,Wang2019,Wang2021,Banados2018,Banados2021,Matsuoka2018,Matsuoka2019,Matsuoka2022,Reed2019,Yang2019_discovery,Yang2020,Yang2021}).
These unobscured active galactic nuclei (AGN) are powered by the earliest known supermassive black holes with masses of $M_\mathrm{BH} \approx (0.3-3)\times10^9\ M_\odot$ accreting near the Eddington limit \citep[e.g.][]{Yang2021}.
Far-IR follow-up of these quasars has provided a window into their host galaxies, revealing that several of these systems show very bright [CII], CO, and dust continuum emission similar to sub-mm galaxies \citep[e.g.][]{Venemans2016,Venemans2017_z7molecular,Venemans2017_z7p5,Venemans2020,Decarli2018,Decarli2022,Yang2019_lensedz6p5,Pensabene2021}.
One population that would help bridge the gap between known quasar and sub-mm galaxies at $z\gtrsim7$ are heavily-obscured AGN.
However, even though these obscured AGN are expected to be more abundant than their unobscured counterparts (i.e. quasars) in the early Universe \citep{Vito2018_obscuredFraction,Davies2019,Trebitsch2019,Ni2020,Lupi2022}, to date no such object has been confirmed at $z>6.5$.

We previously reported the photometric identification of a heavily-obscured, hyperluminous AGN candidate at $z\simeq6.83$ located in the 1.5 deg$^2$ COSMOS field (\citealt{Endsley2022_radioAGN}, hereafter \citetalias{Endsley2022_radioAGN}). 
This source, COS-87259, was originally identified as a Lyman-break $z\simeq6.6-6.9$ galaxy using deep optical$+$near-IR imaging \citep{Endsley2021_OIII}, and was later found to exhibit a sharp spectral discontinuity between two intermediate/narrow-bands at 9450 \AA{} and 9700 \AA{} suggesting $\zphot{} = 6.83\pm0.06$ (\citetalias{Endsley2022_radioAGN}).
While similar in UV luminosity to many other wide-area selected Lyman-break $z\sim7$ galaxies ($J$ = 25.0 AB; $\Muv{} = -21.7$), COS-87259 was found to show both an exceptionally red rest-UV slope ($\beta = -0.6$) and a prominent Balmer break in the \textit{Spitzer}/IRAC bands, implying unusually strong dust attenuation as well as a relatively old, massive stellar population \citep{Endsley2021_OIII}.
Subsequent investigations revealed that this galaxy is also detected in \textit{Spitzer}/MIPS, \textit{Herschel}/PACS, \textit{Herschel}/SPIRE, JCMT/SCUBA-2, VLA, LOFAR, and MeerKAT bands (\citetalias{Endsley2022_radioAGN}).
These data imply that, if at $z\simeq6.8$, COS-87259 is undergoing rapid, highly-obscured star formation activity and harbors a heavily-obscured hyperluminous radio-loud ($L_{1.4\ \mathrm{GHz}}=10^{25.4}$ W Hz$^{-1}$) AGN (\citetalias{Endsley2022_radioAGN}).

Here, we report the ALMA spectroscopic confirmation of COS-87259 as well as detections of exceptionally strong [CII]158$\mu$m and underlying dust continuum emission from this system.
We describe the ALMA observations in \S\ref{sec:observations} and then report the [CII] and dust continuum properties of COS-87259 in \S\ref{sec:results} comparing to other known objects at $z>6.5$.
We then discuss the physical properties of COS-87259 in \S\ref{sec:physicalProperties}. 
Our conclusions are summarized in \S\ref{sec:summary}, followed by a brief outlook on the potential of future \textit{Roman} and \textit{Euclid} surveys to identify many more extremely red, UV-bright $z\gtrsim7$ galaxies similar to COS-87259.
Throughout this paper, we quote magnitudes in the AB system \citep{OkeGunn1983}, adopt a flat $\Lambda$CDM cosmology with parameters $h = 0.7$, $\Omega_\mathrm{M} = 0.3$, and $\Omega_\mathrm{\Lambda} = 0.7$, and report 68\% confidence interval uncertainties unless otherwise specified.

\begin{figure}
\includegraphics{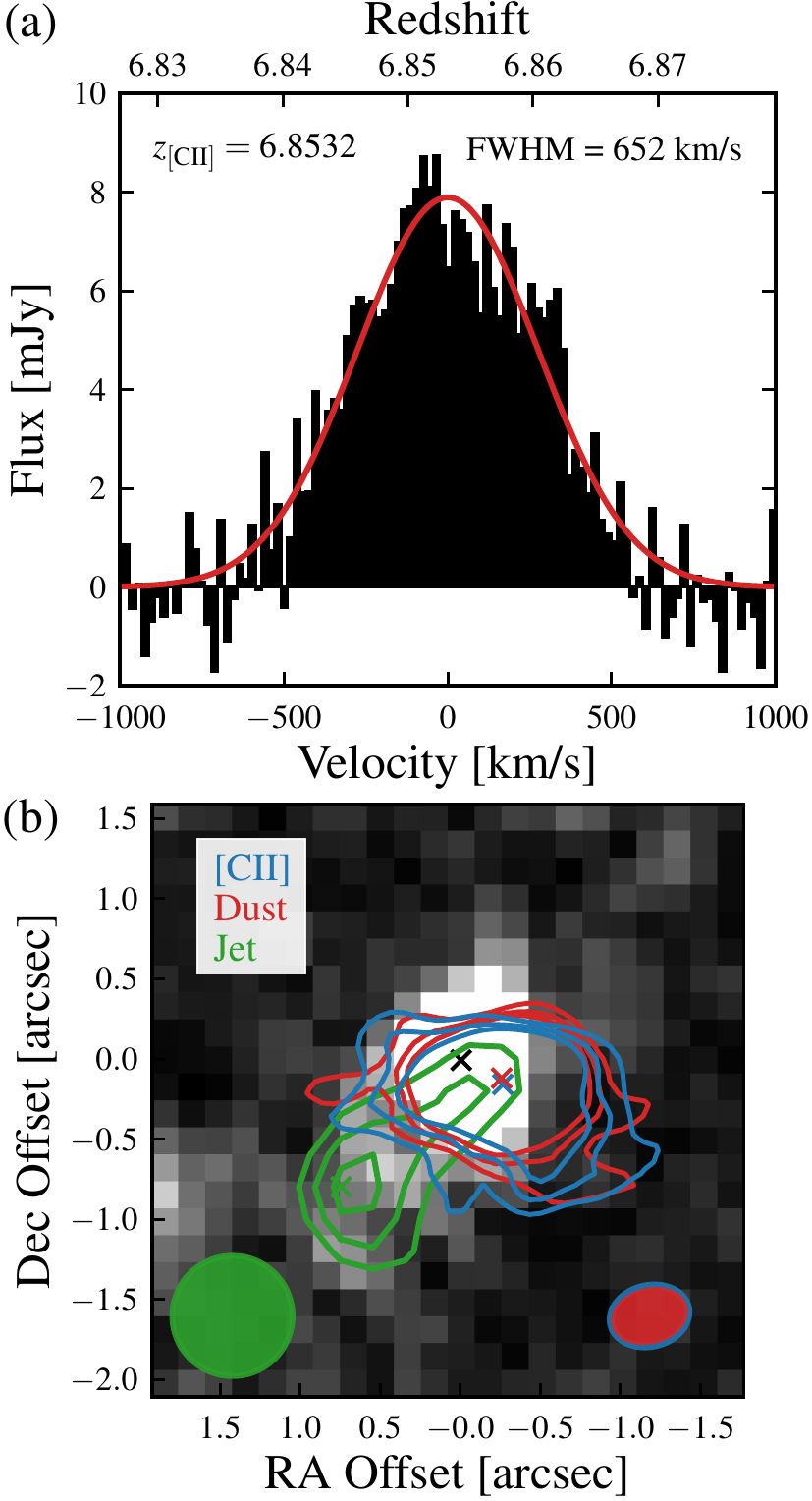}
\caption{\textbf{(a)} ALMA [CII]158$\mu$m spectrum of COS-87259 (after continuum subtraction) extracted within the 3$\sigma$ contours of the moment-0 map. The spectrum is binned to half the original resolution for clarity and the red line shows the Gaussian fit. \textbf{(b)} ALMA [CII] (blue), 158$\mu$m dust continuum (red), and VLA 3 GHz (green) 2, 3, and 4$\sigma$ contour maps of COS-87259 overlaid on the stacked rest-UV image. The ALMA and VLA beams are shown in the bottom corners with their respective colors. The 3 GHz detection peak (green cross) is clearly offset ($\approx$6 kpc projected) from the [CII] and dust continuum centroids (blue and red crosses), as expected if the radio emission is powered by an extended relativistic jet from the AGN. The centroid of the rest-UV emission (black cross) is also slightly offset from that of the far-IR ($\approx$1.5 kpc), consistent with strong dust opacity variations across the galaxy.}
\label{fig:[CII]andContours}
\end{figure}

\section{Observations} \label{sec:observations}

ALMA observations were conducted targeting the [CII]158$\mu$m emission line from COS-87259 on January 6, 2022 (project 2021.1.01311.S).
Given the precise photometric redshift estimate of COS-87259 ($\zphot{} = 6.83\pm0.06$; \citetalias{Endsley2022_radioAGN}), we used the band 6 receiver to target a frequency range of 239.42--244.79 GHz corresponding to \zCII{} = 6.76--6.94.
During the ALMA observations, a total of 45 antennae were used in a configuration with baseline lengths spanning 15--977 meters.
The total integration time on COS-87259 was 33 minutes with an average precipitable water vapor level of 1.1 mm. 
The source J0948+0022 was used for calibration.
We use the reduced data supplied by the observatory which were processed using CASA \citep{McMullin2007_CASA} v.6.2.1.7 with pipeline v.2021.2.0.128.
The resulting naturally-weighted continuum-subtracted spectra have an rms sensitivity of $\approx$440 $\mu$Jy/beam in 9.7 km/s channels, with a beam size of 0.49 arcsec by 0.37 arcsec and a position angle of $-76.6^\circ$. 
The output continuum image ($\nu = 242.14$ GHz) has very similar beam properties with a size of 0.49 arcsec by 0.35 arcsec and a position angle of $-75.1^\circ$, and reaches an rms sensitivity of 22 $\mu$Jy/beam.

\begin{table}
\centering
\caption{Measured and Derived Properties of COS-87259. The reported [CII] and dust continuum sizes are the deconvolved values.} \label{tab:properties}
\begin{threeparttable}[t]
\begin{tabular}{L{3.7cm}P{4.0cm}}
\hline
\hline
\multicolumn{2}{c}{Rest-UV and radio properties (see \citetalias{Endsley2022_radioAGN})} \Tstrut{} \\[2pt]
UV right ascension & 09:58:58.27 \\[2pt]
UV declination & $+$01:39:20.2 \\[2pt]
$\zphot{}$ & 6.83$\pm$0.06 \\[2pt]
Rest-UV slope & $-$0.59$\pm$0.27 \\[2pt]
$M_{1450}$ & $-$21.7$\pm$0.1 \\[2pt]
1.32--3 GHz radio slope & $-1.57^{+0.22}_{-0.21}$ \\[2pt]
0.144--1.32 GHz radio slope &  $-0.86^{+0.22}_{-0.16}$ \\[2pt]
1.4 GHz luminosity & (2.5$\pm$0.5)$\times$10$^{25}$ W Hz$^{-1}$ \\[2pt]
\hline
\multicolumn{2}{c}{ALMA results} \Tstrut{} \\[2pt]
\zCII{} & 6.8532$\pm$0.0002 \\[2pt]
[CII] FWHM & 652$\pm$17 km s$^{-1}$ \\[2pt]
[CII] size (core) & (1.9$\pm$0.4 kpc) $\times$ ($<$1.0 kpc\tnote{a}\, ) \\[2pt]
[CII] size (extended) & (9.6$\pm$1.1 kpc) $\times$ (4.9$\pm$0.6 kpc) \\[2pt]
$F_\mathrm{[CII]}$ (core) & 1.97$\pm$0.16 Jy km s$^{-1}$ \\[2pt]
$F_\mathrm{[CII]}$ (extended) & 5.76$\pm$0.64 Jy km s$^{-1}$ \\[2pt]
Total [CII] luminosity & (8.82$\pm$0.75)$\times$10$^9$ $L_\odot$ \\[2pt]
Dust continuum size (core) & (1.4$\pm$0.2 kpc) $\times$ (1.0$\pm$0.1 kpc) \\[2pt]
Dust continuum size (extended) & (7.8$\pm$10 kpc) $\times$ (3.4$\pm$0.5 kpc) \\[2pt]
$F_{242\ \mathrm{GHz}}$ (core) & 1.48$\pm$0.05 mJy \\[2pt]
$F_{242\ \mathrm{GHz}}$ (extended) & 1.23$\pm$0.15 mJy \\[2pt]
Total $F_{242\ \mathrm{GHz}}$ & 2.71$\pm$0.16 mJy \\[2pt]
$L_\mathrm{IR,SF}$\tnote{b} & (8.7$\pm$0.5)$\times$10$^{12}\ L_\odot$ \\[2pt]
$M_\mathrm{dust}$\tnote{b} & (1.9$\pm$0.1)$\times$10$^9$ $M_\odot$ \\[2pt]
SFR$_\mathrm{IR}$\tnote{c} & 1300$\pm$70 $M_\odot$ yr$^{-1}$ \\[2pt]
SFR$_\mathrm{[CII]}$\tnote{d} & 2970$\pm$250 $M_\odot$ yr$^{-1}$ \\[2pt]
\hline
\multicolumn{2}{c}{SED fitting results} \Tstrut{} \\[2pt]
$L_\mathrm{IR,AGN}$ & $(2.5\pm0.2)\times10^{13}\ L_\odot$ \\[2pt]
$\tau_{_{\mathrm{9.7\mu m}}}$ & 2.3$\pm$0.1 \\[2pt]
$L_\mathrm{bol,AGN}$ & $(5.1\pm0.5)\times10^{13}\ L_\odot$ \\[2pt]
$M_{1450,\mathrm{unobscured}}$ & $-27.3 \pm 0.1$ \\[2pt]
$M_\mathrm{BH}\, \lambda_\mathrm{Edd}$ & (1.6$\pm$0.2)$\times$10$^9$ $M_\odot$ \\[2pt]
$L_\mathrm{IR,SF}$ & (9.1$\pm$1.1)$\times$10$^{12}$ $L_\odot$ \\[2pt]
SFR & 1610$\pm$190 $M_\odot$ yr$^{-1}$ \\[2pt]
$M_\ast$ & (1.7$\pm$0.7)$\times$10$^{11}$ $M_\odot$ \\[2pt]
\hline
\end{tabular}
\begin{tablenotes}
\item[a] 2$\sigma$ upper limit on minor axis size from \textsc{imfit}.
\item[b] Computed from the 158$\mu$m dust continuum measurement adopting $T_\mathrm{dust} = 47$ K and $\beta_\mathrm{dust} = 1.6$.
\item[c] Converted from $L_\mathrm{IR,SF}$ using the relation of \citet{Murphy2011}.
\item[d] Converted from $L_\mathrm{[CII]}$ using the relation of \citet{HerreraCamus2015}.
\end{tablenotes}
\end{threeparttable}
\end{table}

\section{Results} \label{sec:results}

In this section, we begin by reporting the [CII]158$\mu$m and underlying dust continuum data for COS-87259 (\S\ref{sec:CIIandDust}), summarizing the measured and derived properties in Table \ref{tab:properties}. We then compare the observed far-IR properties of COS-87259 to other known systems at $z>6.5$ (\S\ref{sec:comparison}).
Finally, we quantify the relative positions of the far-IR, rest-UV, and radio emission, and subsequently investigate whether there is any evidence that source confusion is significantly impacting the mid-IR detections of COS-87259 (\S\ref{sec:relativePositions}).

\subsection{[CII]158\texorpdfstring{$\mathbf{\mu}$m}{micron} and Dust Continuum Emission} \label{sec:CIIandDust}

The primary goal of our ALMA observations is to test the high-redshift ($z\simeq6.83$) nature of COS-87259, as implied by the optical through sub-mm photometric SED (\citetalias{Endsley2022_radioAGN}). 
We visually inspect the spectral data cubes, finding a clear emission line detection coincident with the near-IR position of COS-87259.
To generate a 1D spectrum of this emission line, we create a moment-0 map for the extraction.
We determine the frequency range over which to make the moment-0 map by first extracting the (continuum-subtracted) spectra over a single beam centered on the rest-UV emission of COS-87259 and subsequently fitting a Gaussian profile to determine the line FWHM and central frequency.
Following \citet{Schouws2022_CII}, a moment-0 map is then generated by summing all channels that fall within $\pm$FWHM determined above, and we next extract the spectrum within the 3$\sigma$ contours and refit a Gaussian profile to determine the line FWHM and central frequency.
This process is performed iteratively until the central frequency and FWHM converge (within a few iterations) which we then use to create the final moment-0 map and 1D extracted spectrum.

From a Gaussian fit to the final 1D spectrum (Fig. \ref{fig:[CII]andContours}a), we measure a FWHM = 652$\pm$17 km s$^{-1}$ and a central frequency of 242.009$\pm$0.007 GHz.
This central frequency corresponds to a [CII] redshift of $\zCII{} = 6.8532\pm0.0002$, in excellent agreement with the precise photometric redshift of $\zphot{} = 6.83\pm0.06$ set by the sharp spectral break seen between the Subaru/Hyper Suprime-Cam intermediate/narrow-bands at 9450 \AA{} and 9700 \AA{} (\citetalias{Endsley2022_radioAGN}).
We therefore conclude that the ALMA-detected emission feature seen from COS-87259 is the [CII]158$\mu$m line. 

\begin{figure*}
\includegraphics{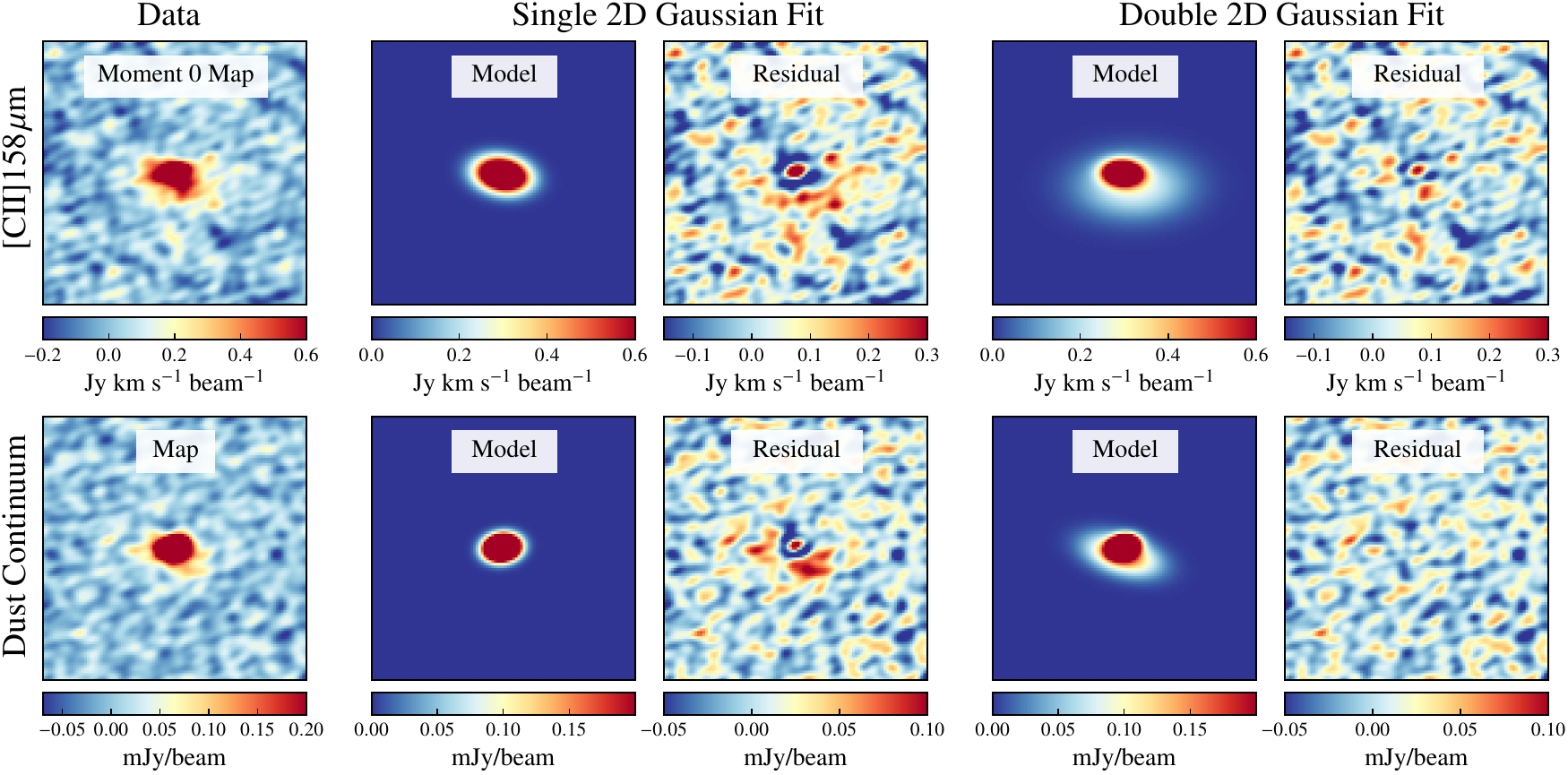}
\caption{Illustration of the 2D Guassian fits to the [CII] moment 0 map (top) and the 158$\mu$m dust continuum map (bottom). The data are shown in the leftmost panels, with the best-fitting single elliptical Gaussian models shown in the second panels from left. These single Gaussian fits provide a poor match to the measured surface brightness profiles of both the [CII] and dust continuum data, as can be seen by the strong residuals in the middle panels. The data are much better fit as the superposition of two elliptical Gaussians as illustrated in the two rightmost panels. Each panel is 7.3 arcsec (39 kpc) on a side.}
\label{fig:2DgaussianFits}
\end{figure*}

\begin{figure}
\includegraphics{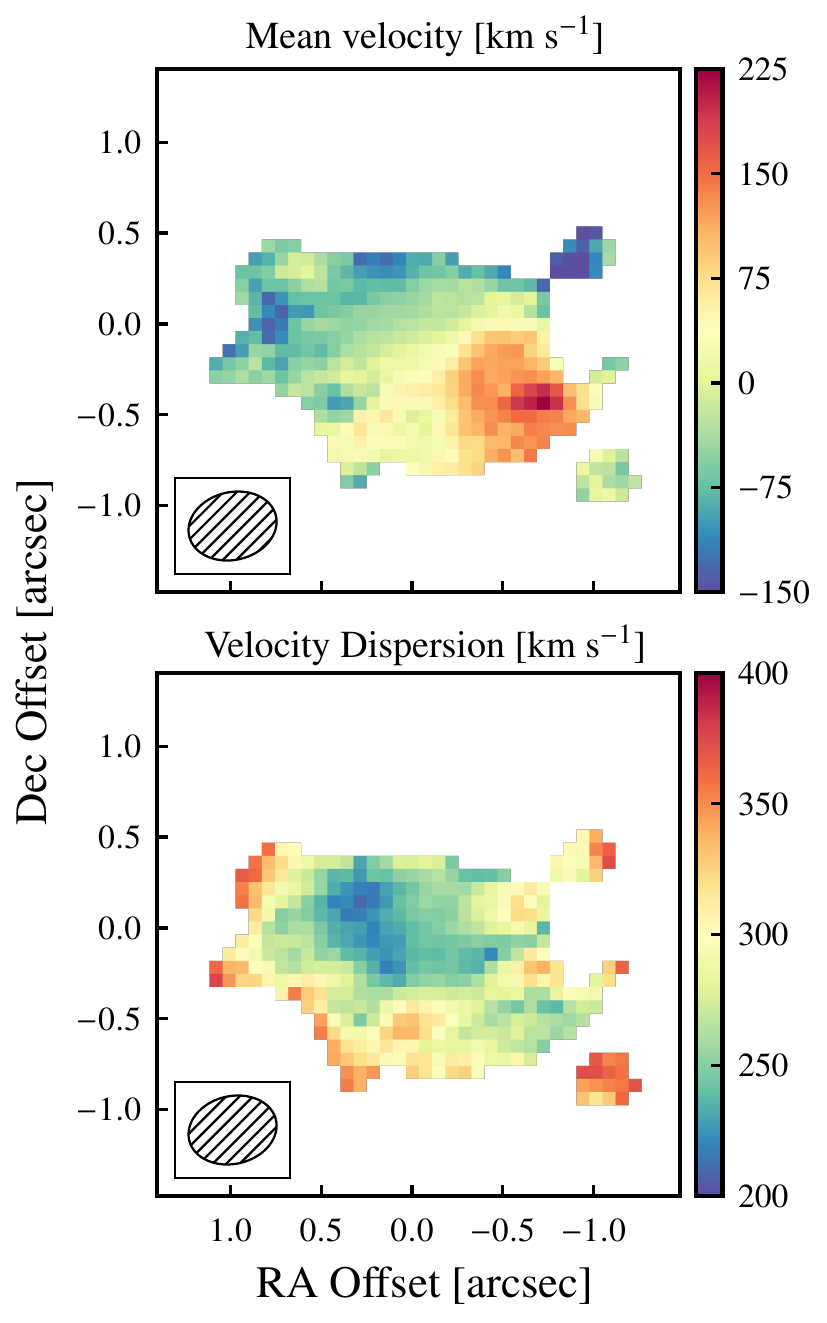}
\caption{Maps of the [CII] mean velocity (top) and velocity dispersion (bottom) fields of COS-87259 derived using \textsc{qubefit}. The footprint of each map is confined to pixels with $>$3 sigma [CII] flux and the [CII] beam ($\approx$0.4 arcsec) is shown in the bottom left corner of each panel.  A prominent gradient is seen in the mean velocity field and there is evidence of larger velocity dispersion towards the outer edges of the [CII] emission.}
\label{fig:momentMaps}
\end{figure}

The [CII] emission of COS-87259 is spatially resolved with our moderate-resolution (beam$\approx$0.4 arcsec) ALMA data (Fig. \ref{fig:[CII]andContours}b).
To determine the flux and size of the [CII] emission, we use the CASA task \textsc{imfit} which fits one or more 2D Gaussians to the moment-0 map.
We find that a single elliptical Gaussian poorly reproduces the observed [CII] surface brightness profile, while a two-component model provides a satisfactory match to the data (Fig. \ref{fig:2DgaussianFits}).
This two-component fit suggests the presence of a compact ($\approx$2 kpc) [CII]-emitting core in addition to a more extended ($\approx$10 kpc) region of emission, consistent with the [CII] surface brightness profiles of $z>6$ quasar hosts \citep{Novak2020}.
The extended [CII] component of COS-87259 has a deconvolved size of (1.82$\pm$0.21 arcsec)$\times$(0.93$\pm$0.11 arcsec) and an integrated flux of 5.76$\pm$0.64 Jy km s$^{-1}$.
At the [CII] redshift of COS-87259 ($\zCII{} = 6.853$), these values translate to a physical size and [CII] line luminosity of (9.6$\pm$1.1 kpc)$\times$(4.9$\pm$0.6 kpc) and (6.57$\pm$0.73)$\times$10$^{9}$ $L_\odot$, respectively.
The compact component is only resolved along the major axis with an estimated deconvolved size of 0.35$\pm$0.07 arcsec (1.9$\pm$0.4 kpc) and its integrated flux is found to be 1.97$\pm$0.16 Jy km s$^{-1}$, which corresponds to a luminosity of (2.25$\pm$0.18)$\times$10$^{9}$ $L_\odot$.
The two-component \textsc{imfit} output therefore implies a total [CII] luminosity of (8.82$\pm$0.75)$\times$10$^{9}$ $L_\odot$, the bulk of which is originating from the extended component (75$\pm$10\%).

The [CII] emission line is frequently used as a valuable tracer of the ISM kinematics for reionization-era galaxies \citep[e.g.][]{Smit2018,Hashimoto2019,Matthee2019_resolvedUVandCII,Neeleman2019}.
We begin investigating the kinematics of COS-87259 by generating maps of the [CII] mean velocity and velocity dispersion fields using the code \textsc{qubefit} \citep{Neeleman2021}.
When running \textsc{qubefit}, we apply a 1$\sigma$ clipping as a compromise between mitigating the impact of noise on the resulting moment maps while preserving true signal (see discussion in \citealt{Neeleman2021}). We have verified that our conclusions do not significant change if we instead adopt a clipping threshold in the range 0.5--2$\sigma$.
A prominent gradient is found in the mean velocity field with a maximum difference of $\approx$360 km s$^{-1}$ between spaxels in the northeast and southwest corners (Fig. \ref{fig:momentMaps}). 
The velocity dispersion map also shows that the outermost spaxels tend to have a larger fitted FWHM ($\sim$600--900 km s$^{-1}$) than those located near the peak of emission ($\approx$550 km s$^{-1}$; see Fig. \ref{fig:momentMaps}). 
This perhaps suggests the presence of a broader [CII] wing in COS-87259 as reported in other $z>6$ AGN hosts \citep{Maiolino2012,Izumi2021} and stacked spectra of $z=4-6$ galaxies \citep{Ginolfi2020}, implying significant outflowing gas. 
However, because the [CII] FWHM only varies by a factor of $\approx$1.5 across COS-87259 (c.f. the factor of 4 in \citealt{Izumi2021}), it is much more challenging to determine if a broad wing is indeed present in this system with existing moderate-resolution data. 
Upon extracting the [CII] spectrum across the outermost spaxels in the moment-0 footprint, we find that the resulting profile has a slightly larger fitted FWHM ($690^{+60}_{-50}$ km/s), though consistent with the full profile shown in Fig. \ref{fig:[CII]andContours}a within uncertainties.
We also extract position-velocity information along the major axis (as determined by a single 2D Gaussian fit) and find no clear evidence of multiple distinct components separated in velocity space (see Fig. \ref{fig:PV}).
Higher-resolution [CII] data will be required to better assess whether the kinematics of COS-87259 are more consistent with a rotating disk or a disturbed, possibly merging system. 

Strong dust continuum emission is also detected from COS-87259 (Fig. \ref{fig:[CII]andContours}b).
Similar to the [CII] emission, we find that the dust continuum surface brightness profile is much better fit by the superposition of two elliptical Gaussians as opposed to a single Gaussian (Fig. \ref{fig:2DgaussianFits}). 
With \textsc{imfit}, we find a deconvolved angular size of (1.47$\pm$0.19 arcsec)$\times$(0.65$\pm$0.10 arcsec) and (0.26$\pm$0.03 arcsec)$\times$(0.18$\pm$0.02 arcsec) for the more extended and compact components, respectively.
These scales correspond to physical sizes of (7.8$\pm$1.0 kpc)$\times$(3.4$\pm$0.5 kpc) and (1.4$\pm$0.2 kpc)$\times$(1.0$\pm$0.1 kpc), respectively, which are systematically $\approx$20--30\% smaller than that found for the [CII] emission.
The integrated flux densities of the more compact and extended components are nearly equal at 1.48$\pm$0.05 mJy and 1.23$\pm$0.15 mJy, respectively, yielding a total 158$\mu$m dust continuum flux density of 2.71$\pm$0.16 mJy for COS-87259.

Given the highly luminous nature of COS-87259, we investigate whether this system may be significantly magnified by nearby foreground objects. 
We consider all foreground objects within 10 arcsec of COS-87259 identified in the COSMOS2020 catalog \citep{Weaver2022_COSMOS2020}, assuming each is a singular isothermal sphere with photometric redshift and stellar mass taken from that same catalog.
The velocity dispersion of each neighboring source is computed in two independent ways.
First, we assume that each galaxy occupies a virialized halo with 200$\times$ the mean density of the Universe at its assumed redshift and mass 100$\times$ that of the stellar mass.
Our second approach follows that of \citet{Mason2015} where we adopt the relation between velocity dispersion and stellar mass from \citet{Auger2010}.
Both methods to estimate velocity dispersions suggest that COS-87259 is only weakly lensed by neighboring foreground sources, with a small combined magnification factor of $\mu_\mathrm{tot} \approx 1.15 - 1.25$.
Because this weak magnification does not significantly impact our main conclusions, we choose not to correct for it when reporting the measured and inferred properties of COS-87259.

\begin{figure}
\includegraphics{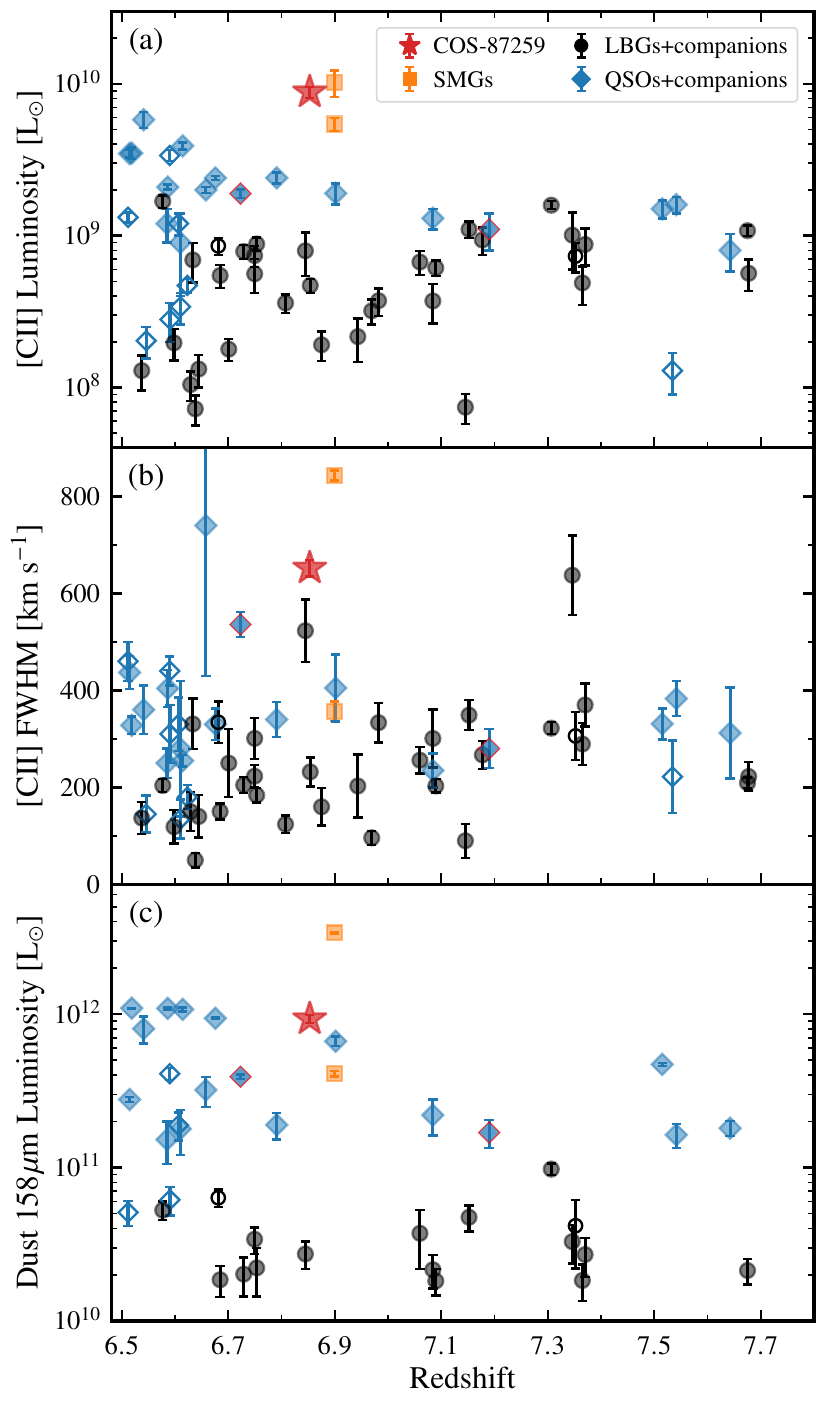}
\caption{Comparison of the observed far-infrared properties of COS-87259 with other known objects at $z>6.5$. We compare the [CII] luminosity, [CII] FWHM, and monochromatic 158$\mu$m dust continuum luminosities in panels (a), (b), and (c), respectively. A full description of the objects shown in this plot is provided in \S\ref{sec:comparison}. The two known red type 1 quasars at $z>6.5$ are shown as blue diamonds with red edges. Serendipitously-detected companions of quasars and Lyman-break galaxies are shown as empty markers. We correct for strong gravitational lensing where appropriate.}
\label{fig:literatureComparison}
\end{figure}

\subsection{Comparison of Far-IR Properties to Other \texorpdfstring{$\mathbf{z>6.5}$}{z>6.5} Sources} \label{sec:comparison}

We compare the observed far-infrared properties ([CII] luminosity, [CII] FWHM, and monochromatic 158$\mu$m dust continuum luminosity) of COS-87259 to other known objects in the reionization-era.
From the literature, we compile such measured properties from known $z>6.5$ sub-mm galaxies (\citealt{Strandet2017,Marrone2018}), extremely UV-luminous quasars ($\Muv{} \leq -25$;  
\citealt{Venemans2012,Banados2015,Venemans2016,Venemans2017_z7p5,Mazzucchelli2017,Decarli2018,Yang2019_lensedz6p5,Yang2020,Wang2021,Yue2021}), UV-bright Lyman-break galaxies ($\Muv{} \leq -21$; \citealt{Pentericci2016,Smit2018,Hashimoto2019,Bouwens2022_REBELS,Inami2022,Schouws2022_dust,Schouws2022_CII}), reddened type 1 quasars \citep{Izumi2021,Fujimoto2022}, as well as companion galaxies serendipitously identified from ALMA observations \citep{Decarli2017,Willott2017,Neeleman2019,Venemans2019,Venemans2020,Fudamoto2021}.
We correct the properties of the sub-mm system SPT0311-58 and the quasar J0439+1634 for strong gravitational lensing given reported magnitification factors \citep{Marrone2018,Yue2021}.

With a [CII] luminosity of (8.82$\pm$0.75)$\times$10$^{9}$ $L_\odot$, COS-87259 is the second most [CII] luminous object known at $z>6.5$ (Fig. \ref{fig:literatureComparison}a), being outshined by only the sub-mm galaxy SPT0311-58W at $z=6.900$ (1.02$\times$10$^{10}$ $L_\odot$).
The [CII] line of COS-87259 is nearly 5$\times$ brighter than typical extremely bright $z>6.5$ quasars (median L$_\mathrm{[CII]} = 2.0\times10^9\ L_\odot$) and 16$\times$ brighter than typical UV-bright Lyman-break $z>6.5$ galaxies (median L$_\mathrm{[CII]} = 0.55\times10^9\ L_\odot$).
COS-87259 additionally shows a [CII] line that is 4.7--8.0$\times$ brighter than the two reddened type 1 quasars thus far identified at $z>6.5$ \citep{Izumi2021,Fujimoto2022}.

The [CII] line width of COS-87259 (FWHM=652$\pm$17 km s$^{-1}$) is also exceptionally broad for known systems at $z>6.5$ (Fig. \ref{fig:literatureComparison}b), only surpassed by SPT0311-58W (FWHM = 840$\pm$10 km s$^{-1}$; \citealt{Marrone2018}) and the $z=6.66$ quasar PSO J338+29 (FWHM = 740$^{+540}_{-310}$ km s$^{-1}$; \citealt{Mazzucchelli2017}).
For reference, the typical [CII] line widths of extremely UV-luminous $z>6.5$ quasars and Lyman-break galaxies are 2.0 and 2.7$\times$ narrower than that of COS-87259, respectively.
Finally, we note that the monochromatic 158$\mu$m dust continuum luminosity of COS-87259 ($(9.3\pm0.5) \times 10^{11}\ L_\odot$) is among the brightest of any known object at $z>6.5$ (Fig. \ref{fig:literatureComparison}c).
This 158$\mu$m continuum luminosity is very similar to that of $z>6.5$ quasars, with only SPT0311-58W substantially outshining COS-87259 in the continuum around [CII].
While the conversion between 158$\mu$m luminosity to total infrared luminosity depends on dust temperature and emissivity, it is nonetheless clear that COS-87259 is among the most extreme objects known in the reionization era in terms of its observed [CII] and dust continuum properties.
This suggests that COS-87259 is likely one of the most massive galaxies at $z\sim7$ with a very large star formation rate, as supported by the full near-infrared to millimeter SED of this system (see \S\ref{sec:physicalProperties}).

\begin{figure}
\includegraphics{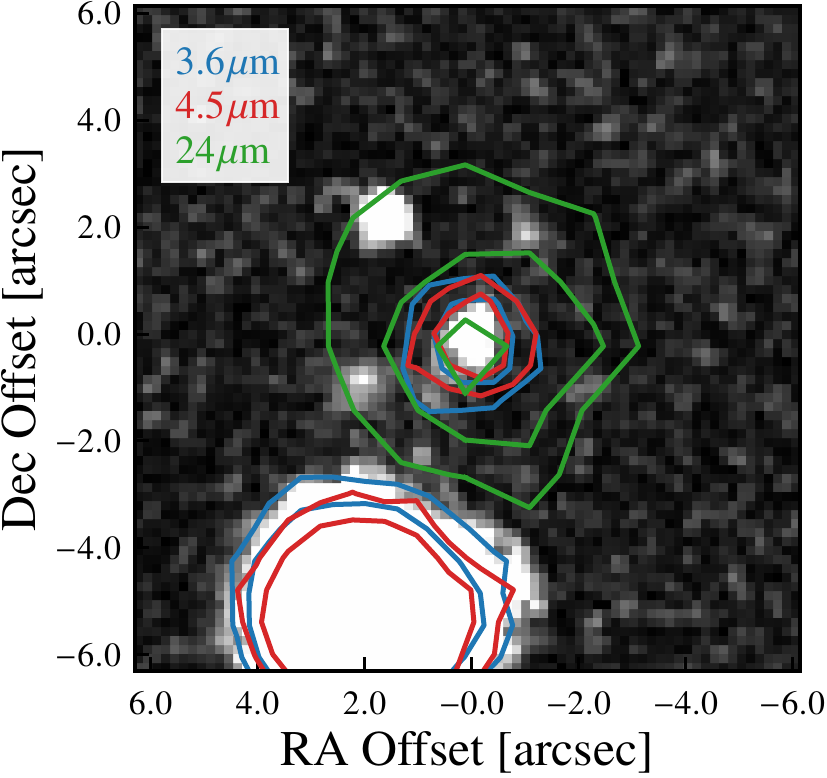}
\caption{Figure showing the \textit{Spitzer}/IRAC 3.6$\mu$m (blue), 4.5$\mu$m (red), as well as \textit{Spitzer}/MIPS 24$\mu$m (green) emission around COS-87259. We show 7 and 12$\sigma$ contours for IRAC, and 4, 6, and 8$\sigma$ contours for MIPS. The background image is the near-IR stack centered on the rest-UV position of COS-87259. The centroids of each \textit{Spitzer} detection near COS-87259 are closely aligned with the rest-UV centroid ($<$0.5 arcsec) and there is no evidence of significant confusion impacting any of the \textit{Spitzer} detections.}
\label{fig:IRACmipsContours}
\end{figure}

\subsection{Relative Positions of the Multi-wavelength Emission} \label{sec:relativePositions}

To further characterize COS-87259, we compare the relative positions of the [CII], dust continuum, rest-UV, and radio emission, all of which are imaged with FWHM$\approx$0.4--0.8 arcsec ($\approx$2--4 kpc) resolution.
The centroids of the [CII] and dust continuum emission are consistent within uncertainties ($\Delta \theta = 0.04\pm0.03$ arcsec), indicating that the 158$\mu$m dust continuum is concentrated near the dynamical center of the galaxy.
Here, we have computed 1$\sigma$ positional uncertainties from the ALMA data as\footnote{\url{https://help.almascience.org/kb/articles/what-is-the-absolute-astrometric-accuracy-of-alma}} (beam FWHM)/(peak S/N)/0.9, enforcing a minimum positional uncertainty of $\sigma_\mathrm{min}$ = (beam FWHM)/20/0.9 due to limits imposed by atmospheric phase fluctuations. 
We also find that the 3 GHz radio emission peak from COS-87259 (beam=0.75 arcsec; \citealt{Smolcic2017}) is significantly offset from the dust continuum centroid with a separation of 1.21$\pm$0.17 arcsec (see Fig. \ref{fig:[CII]andContours}b). 
This large (6.4$\pm$0.9 kpc projected) offset is to be expected if the radio emission seen from COS-87259 is powered by an extended relativistic jet emerging from the AGN.

The rest-UV emission centroid from COS-87259 also appears slightly offset from both the [CII] and dust continuum centroids (Fig. \ref{fig:[CII]andContours}b). 
We measure the rest-UV centroid from a stack of the HSCSSP PDR2 $y$ and UltraVISTA DR4\footnote{\url{https://ultravista.org/release4/dr4_release.pdf}} $Y$, $J$, $H$, and $K_s$ images, each of which have a seeing-limited FWHM$\approx$0.77 arcsec and are calibrated to the Gaia astrometric frame with an absolute position uncertainty of $\leq$0.05 arcsec \citep{McCracken2012,Aihara2019}.
The stacked rest-UV detection of COS-87259 has a S/N = 15 suggesting a relative positional uncertainty of FWHM/(S/N) $\approx$ 0.05 arcsec, comparable to the absolute precision. 
To be conservative, we adopt a total position uncertainty of 0.10 arcsec for the rest-UV emission.
This results in a total offset of 0.30$\pm$0.10 arcsec between the rest-UV and [CII] emission centroids, as well as a very similar offset from the dust continuum.
Such moderate ($\sim$1--2 kpc) offsets between the dust continuum and rest-UV centroids have been found among other UV-bright Lyman-break $z\sim7$ galaxies and are often coupled with strong spatial gradients in the rest-UV slope, indicating considerable dust opacity variations across the galaxy \citep[e.g.][]{Carniani2018,Bowler2022}.

COS-87259 is also detected at fairly high S/N ($\sim$10--20$\sigma$) in the \textit{Spitzer}/IRAC 3.6$\mu$m, 4.5$\mu$m, and MIPS 24$\mu$m bands (see Fig. \ref{fig:IRACmipsContours}; \citetalias{Endsley2022_radioAGN}).
Given the relatively poor angular resolution of the \textit{Spitzer} images (FWHM$\approx$1.8--6 arcsec), we investigate whether there is any evidence that these detections are significantly impacted by confusion.
After calibrating the \textit{Spitzer} images to the Gaia astrometric frame (using the IRAF task \textsc{ccmap}), we find that the centroids of all detections are closely aligned with the rest-UV centroid of COS-87259 ($\Delta \theta < 0.5$ arcsec).
Moreover, none of the \textit{Spitzer} contours show any sign of strong asymmetry (see Fig. \ref{fig:IRACmipsContours}).
The next nearest MIPS 24$\mu$m detection is 12 arcsec (i.e. 2$\times$ the FWHM) away from COS-87259, while the next nearest strong IRAC detection is from an object 5.4 arcsec away (i.e. 3$\times$ the FWHM) located at the bottom of Fig. \ref{fig:IRACmipsContours}.
Overall, we find no evidence that nearby objects are contributing significantly to the \textit{Spitzer} detections of COS-87259.
Nonetheless, the IRAC, MIPS, PACS, and SPIRE photometry of COS-87259 used below were all computed after performing deconfusion using the relatively high-resolution imaging as a prior (see \citetalias{Endsley2022_radioAGN} for further details).

\section{The Physical Properties of COS-87259} \label{sec:physicalProperties}

Now equipped with measurements of the [CII] systemic redshift and 158$\mu$m dust continuum emission, we reassess the inferred physical properties of COS-87259 derived in \citetalias{Endsley2022_radioAGN}.
To start, we fit the observed near-infrared to millimeter SED of COS-87259 using a custom Bayesian SED fitting package \citep{Lyu2022_GOODSS} that utilizes dynamic nested sampling (\textsc{dynesty}; \citealt{Speagle2020}) and features (semi-)empirical templates of AGN and star-forming dust emission. 
The validity of this tool has been demonstrated by finding and characterizing obscured AGN within the GOODS-S field in \citet{Lyu2022_GOODSS}, and we refer the interested reader to that paper as well as \citet[][their section 4.2]{LyuRieke2022} for details of this code. 
For the purposes of this work, we adopt the \citet{Lyu2017} warm-dust-deficient AGN SED (which is suitable for very luminous AGN like COS-87259; see below) and account for obscuration using an empirical AGN infrared attenuation curve \citep{LyuRieke2022}.
The galaxy dust emission is represented by the infrared SED template of Haro 11 which successfully matches the observed infrared SEDs and expected physical properties of massive $5<z<7$ galaxies \citep{DeRossi2018}, including the hosts of $z>5$ quasars \citep{Lyu2016}. 
Along with these AGN and galaxy dust emission components, we include stellar emission using the Flexible Stellar Population Synthesis code (\textsc{fsps}; \citealt{Conroy2009_FSPS,Conroy2010_FSPS}) under the SED fitting framework provided by \textsc{prospector} \citep{Johnson2021}. 
For this stellar emission, we adopt a delayed star formation history, a \citet{Calzetti2000} attenuation curve (as is likely appropriate for a dust and hence metal rich system like COS-87259; e.g. \citealt{Shivaei2020}), and a \citet{Kroupa2001} stellar initial mass function.

In the SED fitting procedure, we include all VIRCam, WFC3, IRAC, MIPS, PACS, SPIRE, and SCUBA-2 photometry of COS-87259 (see table 1 of \citetalias{Endsley2022_radioAGN}) as well as the 158$\mu$m dust continuum measurement from our new ALMA data.
The only difference from the photometry adopted in \citetalias{Endsley2022_radioAGN} is that we here use the deboosted SCUBA-2 850$\mu$m flux density measurement of 4.60$\pm$1.65 mJy \citep{Simpson2019}.
The IRAC, MIPS, PACS, and SPIRE photometry were all computed using deconfusion algorithms (see \citetalias{Endsley2022_radioAGN}).
We note that the reported uncertainties on inferred physical properties below are largely dominated by the statistical uncertainties from the photometric measurements and do not include systematic uncertainties arising from our adopted set of SED templates.
Using the \textsc{x-cigale} SED-fitting code \citep{Yang2020_xcigale} as described in \citetalias{Endsley2022_radioAGN}, we infer similar values for the physical parameters (within $\lesssim$0.3 dex).

Before reporting the results from our SED fit, we briefly discuss the photometric properties of COS-87259 to provide context on what is driving the inferred physical parameters (see \citetalias{Endsley2022_radioAGN} for further details).
From the VIRCam photometry, COS-87259 shows an extremely red rest-UV slope ($\beta = -0.6$ where $f_\lambda \propto \lambda^\beta$) compared to typical UV-bright $z\sim7$ Lyman-break galaxies ($\beta \approx -1.8$; \citealt{Bowler2017,Endsley2021_OIII}), indicating unusually strong dust attenuation.
At slightly redder wavelengths, COS-87259 shows a jump in flux density between the $K_s$ ($0.67\pm0.09$ $\mu$Jy) and IRAC 3.6$+$4.5$\mu$m bands ($\approx$2.75 $\mu$Jy) that is much stronger than expected from extrapolating the rest-UV slope, consistent with a prominent Balmer break. 
The flux density then steeply rises between the IRAC 8$\mu$m ($<$4.9 $\mu$Jy 2$\sigma$), MIPS 24$\mu$m (180$\pm$6 $\mu$Jy), and PACS 100$\mu$m (5100$\pm$1400 $\mu$Jy) bands, as would be expected from hot dust emission powered by a highly-obscured AGN.
Finally, we note that COS-87259 shows a relatively flat and bright SED between the PACS, SPIRE, and SCUBA-2 160--850$\mu$m bands ($\approx$5--10 mJy in each filter), indicative of very luminous hot (AGN powered) and warm (star formation powered) dust emission, the latter of which is now corroborated by the strong ALMA 158$\mu$m continuum detection.
As discussed in \citetalias{Endsley2022_radioAGN}, the MIPS, PACS, SPIRE, and SCUBA-2 photometry of COS-87259 are consistent with the SEDs of hot dust-obscured galaxies identified at lower redshifts (e.g. \citealt{Eisenhardt2012,Tsai2015,Fan2016_hotDOG}).

\begin{figure}
\includegraphics{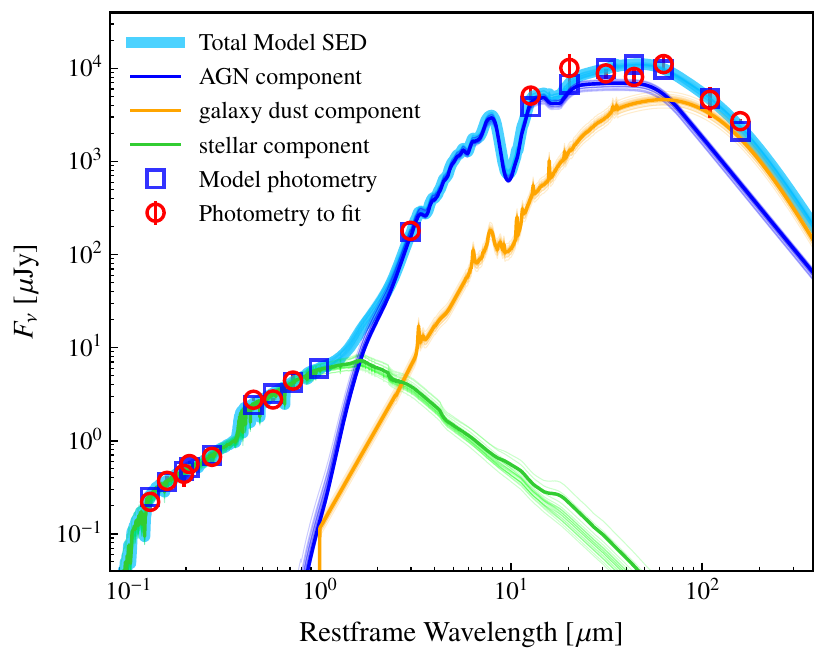}
\caption{Near-infrared through millimeter SED of COS-87259. The photometric data (see \citetalias{Endsley2022_radioAGN}) and ALMA band 6 continuum measurement are shown as red circles with the best-fitting SED and its corresponding model photometry shown via a thick, light blue line and blue squares, respectively. We also show the relative contributions of the stellar, AGN, and (star-forming) dust emission as green, blue, and gold lines, respectively. The transparent lines show 20 realizations of the fit randomly sampled from the output posterior distribution to illustrate the fitting uncertainties. COS-87259 is inferred to be an extremely massive reionization-era galaxy harboring a heavily-obscured luminous AGN and powering intense dust-obscured star formation.}
\label{fig:SEDfit}
\end{figure}

From the SED fit (see Fig. \ref{fig:SEDfit}), COS-87259 is inferred to harbor a very luminous AGN ($L_\mathrm{IR,AGN} = (2.5\pm0.2)\times10^{13}\ L_\odot$) that is heavily obscured with an optical depth at rest-frame 9.7$\mu$m of $\tau_{_{\mathrm{9.7\mu m}}} = 2.3\pm0.1$.
Such inferred obscuration is consistent with Compton-thick absorption ($N_H \gtrsim 10^{24}$ cm$^{-2}$; \citealt{Shi2006}) and helps explain the weak X-ray luminosity of COS-87259, which remains undetected in the 160 ks \textit{Chandra} imaging over COSMOS (\citealt{Civano2016}; see \citetalias{Endsley2022_radioAGN} for further details).
Using the \citet{Runnoe2012} bolometric conversion from rest-frame 5100 \AA{} luminosity, the intrinsic (de-reddened) AGN SED implied by our fits suggests a bolometric AGN luminosity of $L_\mathrm{bol,AGN} = (5.1\pm0.5)\times10^{13}\ L_\odot$.
Such a large bolometric AGN luminosity is comparable to that of $z\sim1-4$ hot dust obscured galaxies \citep[e.g.][]{Eisenhardt2012,Tsai2015,Fan2016_hotDOG} as well as $z\sim7$ quasars \citep[e.g.][]{Mazzucchelli2017,Yang2021}, and implies a black hole mass of $M_\mathrm{BH} = (1.6\pm0.2)\times10^9 M_\odot/\lambda_\mathrm{Edd}$ for COS-87259 where $\lambda_\mathrm{Edd}$ is the Eddington ratio.
We come back to discuss potential implications of identifying such an early supermassive black hole over a relatively small area (1.5 deg$^2$) at the end of this section.

Due to the strong jump in flux density between the VIRCam $K_s$ and IRAC bands, the rest UV$+$optical emission of COS-87259 is inferred to be powered by star formation (with a prominent Balmer break) rather than by the AGN (Fig. \ref{fig:SEDfit}).
To further assess whether the rest UV$+$optical photometry of COS-87259 may plausibly be due to reddened AGN emission, we also compare these data to empirical spectra of lower-redshift quasars.
Specifically, we fit the VIRCam $Y$, $J$, $H$, and $K_s$ as well as the IRAC 4.5$\mu$m photometry of COS-87259 to the composite $z\sim1.5$ blue quasar spectrum from \citet{Selsing2016} after redshifting to $z=6.853$ and applying a range of V-band optical depths ($0<A_V<2$) using the SMC attenuation law \citep{Pei1992} motivated by previous studies of red quasars \citep[e.g.][]{Richards2003,Hopkins2004}.
In this fit, we do not include the 3.6$\mu$m band as it can be contaminated by strong [OIII]$+$H$\beta$ emission at the redshift of COS-87259.
This fitting procedure results in a best-fitting reduced $\chi^2$ = 11.8 with 3--6$\sigma$ offsets in the $K_s$ and 4.5$\mu$m bands, indicating that the rest UV$+$optical photometry of COS-87259 is poorly explained by such AGN spectra.
Ultimately, UV$+$optical spectra of COS-87259 (taken with e.g. JWST/NIRSpec) will be necessary to determine the physical origin of this emission.

Assuming that the UV$+$optical emission is dominated by star formation, the fiducial SED fit shown in Fig. \ref{fig:SEDfit} returns a total stellar mass of $M_\ast = (1.7\pm0.7)\times10^{11}\ M_\odot$, implying that COS-87259 is an extremely massive $z\sim7$ galaxy as perhaps expected given its extreme far-IR properties (Fig. \ref{fig:literatureComparison}).
However, as discussed elsewhere (\citealt{Mierlo2022,Whitler2023}; see also \citetalias{Endsley2022_radioAGN}), alternative assumptions to the star formation history can yield significantly lower stellar mass estimates for COS-87259 ($M_\ast \sim (1-3)\times10^{10}\ M_\odot$) wherein the strong jump in flux density between $K_s$ and [3.6] is inferred to be strongly assisted by prominent \OIIIHb{} emission.
Nonetheless, the full range of estimated stellar masses for COS-87259 ($\sim$10$^{10-11} M_\odot$) still falls substantially below that expected given local relations between the properties of supermassive black holes and their host galaxies.
With an inferred black hole mass of $\sim$1.5$\times$10$^9$ $M_\odot$, the relation of \citet{HaringRix2004} predicts a stellar mass of $\sim$10$^{12}$ $M_\odot$ for COS-87259.
Similar inferred offsets towards higher black hole mass at fixed galaxy mass have been reported for luminous high-redshift quasars \citep[e.g.][]{Trakhtenbrot2015,Venemans2016,Pensabene2020} as well as hot dust obscured galaxies \citep{Assef2015,Tsai2015}, potentially indicating that the growth of their supermassive black holes outpaced that of the host galaxies at early times.
 
The models also suggest that intense obscured star formation within COS-87259 is powering a total (8--1000$\mu$m rest-frame) IR luminosity of $L_\mathrm{IR,SF} = (9.1\pm1.1)\times10^{12}\ L_\odot$, with a corresponding very large star formation rate of 1610$\pm$190 $M_\odot$/yr. 
For comparison, we also derive the obscured star formation properties of COS-87259 adopting the standard approach for $z>6$ quasars.
In these objects, the dust emission is often assumed to follow a modified blackbody with temperature $T = 47$ K and emissivity index $\beta = 1.6$ \citep[e.g.][]{Beleen2006,Decarli2018,Venemans2020}.
After fixing the SED normalization to match the 158$\mu$m dust continuum measurement of COS-87259, this approach yields an IR luminosity and obscured star formation rate \citep{Murphy2011} of $(8.7\pm0.5)\times10^{12}\ L_\odot$ and $1300\pm70\ M_\odot$/yr, respectively, similar to that obtained from the SED fit using the Haro 11 template.
The modified blackbody fit also implies a very large dust mass of $M_\mathrm{dust} = (1.9\pm0.1)\times10^9$ $M_\odot$.
Finally, we note that if we adopt the SFR-$L_\mathrm{[CII]}$ conversion of \citep{HerreraCamus2015}, we infer a total star formation rate of 2970$\pm$250 $M_\odot$/yr.
While this [CII]-based SFR is 2.3$\times$ larger than that inferred from the dust SED, we note that there remain considerable uncertainties on both the dust temperature and emissivity index for COS-87259.
Regardless, the very bright ALMA detections of COS-87259 make it clear that this object is undergoing vigorous obscured star formation activity comparable to $z\sim7$ sub-mm galaxies and quasar hosts (Fig. \ref{fig:literatureComparison}).
The range of stellar masses ($\sim 10^{10-11} M_\odot$) and SFRs (1300-3000 $M_\odot$/yr) inferred for COS-87259 suggest a specific SFR between 13 to 300 Gyr$^{-1}$, consistent with the range of sSFRs inferred among otherwise typical extremely UV-luminous ($\Muv{} \sim -22$) $z\sim7-8$ galaxies \citep{Topping2022_REBELS}. Additional data will be required to determine whether the AGN within COS-87259 is significantly impacting its star formation efficiency.

With an inferred bolometric AGN luminosity of $\sim$5$\times$10$^{13}$ $L_\odot$, the supermassive black hole within COS-87259 is comparable to that powering the brightest quasars known at $z>6$ \citep[e.g.][]{Yang2021} which are discovered using extremely wide-area ($\gtrsim$1000 deg$^2$) surveys.
The identification of COS-87259 within the relatively small 1.5 deg$^2$ COSMOS field is therefore very surprising.
To better place COS-87259 into context of the UV-luminous quasar population, we extract its de-reddened rest-frame 1450 \AA{} AGN luminosity from the SED fit described in \S\ref{sec:physicalProperties}.
This suggests that COS-87259 is a highly-obscured version of a $M_{1450} \approx -27.3$ quasar which, at $z\sim7$, have been found to be powered by $\approx$1.5$\times$10$^9$ $M_\odot$ black holes accreting near the Eddington limit \citep{Yang2021}.
Such UV-luminous quasars ($M_{1450} \leq -27$) are an exceedingly rare population in the very early Universe with a surface density of $\approx$1 per 3000 deg$^{2}$ at $z=6.6-6.9$ \citep{Wang2019}, the approximate redshift interval over which COS-87259 was selected.
While COS-87259 is only one object and thus may have simply been a fortuitous discovery, we briefly discuss what the spectroscopic confirmation of this object may imply for the assembly of supermassive black holes at $z\gtrsim7$.
We refer the interested reader to section 4.2 of \citetalias{Endsley2022_radioAGN} for a more extended discussion of what COS-87259 may imply for the statistical properties of UV-bright ($\Muv{} \lesssim -21.25$) massive ($M_\ast > 10^{10}$ $M_\odot$) $z\sim7$ galaxies, as well as the possible contribution of this population to the obscured star formation rate density at early times.

The identification of COS-87259 within the 1.5 deg$^2$ COSMOS field versus the very low surface density of similarly luminous $z\sim6.8$ quasars ($\approx$1/3000 deg$^{-2}$; see above) raises the question of whether rest-UV selections may be missing a very large fraction ($\gtrsim$1000$\times$) of heavily-obscured black hole growth in the early Universe.
Observations of both the local Universe and the $z>6$ quasar population provide some insight.
The known population of early UV-luminous quasars suggest that approximately 0.1\% of the total black hole mass density inferred at present day had already formed by $z=6$ \citep[e.g.][]{Marconi2004,Shen2020}.
Therefore, boosting the total black hole mass density at $z>6$ by a factor of $\gtrsim$1000 (to correct for an extremely dominant population of heavily-obscured AGN) would imply little cosmic black hole growth between $0<z<6$.
Such a scenario is inconsistent with the census of AGN activity at these redshifts \citep{Shen2020}.
This line of reasoning suggests it is unlikely that there are $\gtrsim$3 orders of magnitude more buried AGN at $z>6$ than expected based on results from rest-UV selected samples.

However, it remains unknown whether the most massive ($\sim$10$^9$ $M_\odot$) black holes at $z\gtrsim7$ very frequently grew in heavily-obscured phases.
Indirect evidence for such a scenario has emerged in recent years, with spectroscopic observations showing that a large fraction of UV-luminous $z>6$ quasars possess much smaller \Lya{} proximity zones than expected given their black hole masses \citep{Davies2019,Eilers2020,Eilers2021}.
Even though these early supermassive black holes have presumably been growing for several hundred million years \citep{Inayoshi2020}, their proximity zones imply that they have only been ionizing their surrounding IGM for $\sim$10$^6$ years on average \citep{Morey2021}.
This apparent tension can be alleviated if the supermassive black holes powering $z>6$ quasars were heavily obscured for a large fraction ($\gtrsim$95\%) of their lifetime \citep{Eilers2018,Davies2019}.
However, it is not clear what physical mechanism might drive such a very dominant population of heavily-obscured luminous AGN at $z\gtrsim7$ given that X-ray observations imply an obscured ($N_H > 10^{23}$ cm$^{-2}$) fraction of $\sim$80\% among luminous ($L_\mathrm{bol} \sim 10^{13}\ L_\odot$) AGN at $z\sim3-6$ \citep{Vito2018_obscuredFraction}.
One potential contributing factor could be the increasing gas and dust density within massive galaxies towards higher redshifts \citep{Circosta2019}, as suggested by recent hydrodynamic simulations \citep{Trebitsch2019,Ni2020,Lupi2022}.
Ultimately, wider-area searches for very luminous ($L_\mathrm{bol} \gtrsim 10^{13}\ L_\odot$) heavily-obscured AGN at $z>6$ will be required to better assess how common this population is relative to quasars, with subsequent detailed follow-up necessary to characterize the origin of their obscuration.

\section{Summary and Outlook} \label{sec:summary}

In this work, we have reported the results of ALMA follow-up observations targeting the [CII]158$\mu$m line and underlying dust continuum from a previously-reported candidate heavily-obscured radio-loud ($L_{1.4\ \mathrm{GHz}}=10^{25.4}$ W Hz$^{-1}$) AGN at $z\simeq6.83$ identified in the 1.5 deg$^2$ COSMOS field (COS-87259; \citetalias{Endsley2022_radioAGN}).
Our summary and conclusions are listed below:

\begin{enumerate}

    \item The ALMA data confidently reveal a luminous emission line with a central frequency of 242.009$\pm$0.007 GHz coincident with the near-infrared position of COS-87259 (Fig. \ref{fig:[CII]andContours}). We conclude that this emission feature is the [CII]158$\mu$m cooling line at $\zCII{}=6.8532\pm0.0002$ which is in excellent agreement with the precise photometric redshift of COS-87259 ($\zphot{} = 6.83\pm0.06$) set by the sharp Lyman-alpha break seen between two narrow/intermediate-bands at 9450 \AA{} and 9700 \AA{} (\citetalias{Endsley2022_radioAGN}). 
    
    \item The [CII]158$\mu$m emission of COS-87259 is spatially resolved with our moderate-resolution (beam$\approx$0.4 arcsec) ALMA data and its surface brightness profile suggests the presence of a compact (major axis = 1.9$\pm$0.4 kpc) [CII]-emitting core in addition to a more extended (major axis = 9.6$\pm$1.1 kpc) region of emission (Fig. \ref{fig:2DgaussianFits}), similar to that of $z>6$ quasar hosts \citep{Novak2020}.
    Summing the two emission components, we measure a total [CII] line luminosity of (8.82$\pm$0.75)$\times$10$^9$ $L_\odot$ from COS-87259, the bulk of which is originating from the more extended region (75$\pm$10\%). 
    The [CII] emission is extremely broad (FWHM = 652$\pm$17 km s$^{-1}$) and a prominent gradient is seen in the mean velocity field with a maximum difference of $\approx$360 km s$^{-1}$ between spaxels in the northeast versus southwest corners of COS-87259 (Fig. \ref{fig:momentMaps}). 
    
    \item Strong 158$\mu$m dust continuum emission is also detected from COS-87259 with a surface brightness profile consistent with a compact (major axis = 1.4$\pm$0.2 kpc) core and a more extended (major axis = 7.8$\pm$1.0 kpc) region of emission (Fig. \ref{fig:2DgaussianFits}), similar to that of the [CII] line. The combined flux density of these two dust continuum components is measured to be 2.71$\pm$0.16 mJy and is almost equally divided among the two components (55$\pm$4\% from the compact core).
    
    \item COS-87259 exhibits the second most luminous and third broadest [CII] line known at $z>6.5$ and the 158$\mu$m dust continuum monochromatic luminosity of this system is also very similar to that of the IR-brightest $z\sim7$ quasar host galaxies (Fig. \ref{fig:literatureComparison}). The ALMA data therefore indicate that COS-87259 is among the most extreme objects known in the reionization-era in terms of its far-IR properties, suggesting it is likely one of the most massive galaxies at $z\sim7$ with a very large star formation rate.
    
    \item The centroids of the [CII] and dust continuum emission are consistent within uncertainties, though we identify significant offsets between the ALMA centroids and the rest-UV as well as radio centroids. The $\approx$6 kpc projected separation between the 3 GHz detection peak and the [CII]$+$dust continuum centroids is to be expected if the radio emission is powered by an extended relativistic jet emerging from the AGN.
    
    \item The MIPS, PACS, and SPIRE detections of COS-87259 indicate that this source harbors a heavily-obscured AGN with a rest-frame 9.7$\mu$m optical depth of $\tau_{_{\mathrm{9.7\mu m}}} = 2.3\pm0.1$, consistent with Compton-thick absorption. The very large bolometric AGN luminosity of $(5.1\pm0.5)\times10^{13}\ L_\odot$ implies that COS-87259 is a highly-obscured version of a $M_{1450} \approx -27.3$ quasar at $z\sim7$ and is powered by a $\gtrsim$1.5$\times$10$^9$ $M_\odot$ black hole (assuming an Eddington ratio of $\lesssim$1). Comparably luminous quasars are an exceedingly rare population at $z\sim7$ ($\sim$0.001 deg$^{-2}$; \citealt{Wang2019}) while COS-87259 was identified over a relatively small (1.5 deg$^2$) area. Wider-area searches for very luminous ($L_\mathrm{bol} \gtrsim 10^{13}\ L_\odot$) heavily-obscured AGN at $z>6$ will be required to better assess how common this population is relative to quasars.
    
    \item The 158$\mu$m dust continuum emission suggests that COS-87259 is forming stars at a rate of $\approx$1300 $M_\odot$/yr, respectively, consistent with the estimate from the near-IR through millimeter SED fit ($\approx$1610 $M_\odot$/yr). Regardless of the exact value, it is clear from the strong ALMA detections that COS-87259 is undergoing vigorous obscured star formation activity comparable to $z\sim7$ sub-mm galaxies and quasar hosts. The VIRCam and IRAC photometry of COS-87259 suggest that its rest UV$+$optical emission is powered by star formation rather than by the AGN. Under this assumption, we estimate an extremely large stellar mass of (1.7$\pm$0.7)$\times$10$^{11}$ $M_\odot$ for COS-87259, though different assumptions on star formation history can yield significantly lower mass estimates given existing broadband photometry (\citetalias{Endsley2022_radioAGN}; \citealt{Mierlo2022,Whitler2023}). Nonetheless, all stellar mass estimates of COS-87259 are at least an order of magnitude lower than that expected from local scaling relations between the properties of supermassive black holes and their host galaxies. 
    
\end{enumerate}

Future wide-area datasets will provide a greatly improved understanding of the incidence of heavily-obscured supermassive black hole growth and dust-enshrouded stellar mass buildup within the most massive reionization-era galaxies.
Within the coming decade, surveys with \textit{Roman} and \textit{Euclid} are expected to cover hundreds to thousands of square degrees with optical$+$near-infrared (0.9--2$\mu$m) depths of $m_{5\sigma}\sim26-27$ (\citealt{Spergel2015,Euclid2022_deepzgt6,Scaramella2022}). 
These surveys will therefore enable the identification of extremely red ($\beta > -1$) yet UV-bright ($\Muv{} \lesssim -21.5$) $z\gtrsim7$ galaxies like COS-87259 over a much wider area via the presence of very strong ($\gtrsim$3 mag) Lyman-alpha breaks.
Sensitive follow-up at mid/far-infrared, radio, and X-ray wavelengths will provide the opportunity to investigate both obscured star formation as well as heavily-obscured AGN activity within these sources, delivering a much clearer picture of the typical characteristics of the most massive reionization-era galaxies.
Such studies will complement searches of red yet relatively unobscured quasars \citep[e.g.][]{Kato2020,Fujimoto2022}, radio-selected AGN \citep[e.g.][]{Smith2016,Saxena2018,Broderick2022}, and X-ray selected AGN \citep{Nandra2013_Athena,Weisskopf2015_Lynx} in the reionization epoch to build a more complete census of early supermassive black hole growth mechanisms.

\section*{Acknowledgements}

The authors sincerely thank the anonymous referee for providing helpful, constructive comments that significantly improved the quality of this work. RE and DPS acknowledge funding from JWST/near-IRCam contract to the University of Arizona, NAS5-02015. DPS acknowledges support from the National Science Foundation through the grant AST-2109066. FW acknowledges the support provided by NASA through the NASA Hubble Fellowship grant \#HST-HF2-51448.001-A awarded by the Space Telescope Science Institute, which is operated by the Association of Universities for Research in Astronomy, Incorporated, under NASA contract NAS5-26555. JY and XF acknowledge support from the US NSF Grant AST 19-08284. RS acknowledges support from a STFC Ernest Rutherford Fellowship (ST/S004831/1). RJB acknowledges support from TOP grant TOP1.16.057 provided by the Nederlandse Organisatie voor Wetenschappelijk Onderzoek (NWO).

This paper makes use of the following ALMA data: ADS/JAO.ALMA\#2021.1.01311.S. ALMA is a partnership of ESO (representing its member states), NSF (USA) and NINS (Japan), together with NRC (Canada), MOST and ASIAA (Taiwan), and KASI (Republic of Korea), in cooperation with the Republic of Chile. The Joint ALMA Observatory is operated by ESO, AUI/NRAO and NAOJ. The National Radio Astronomy Observatory is a facility of the National Science Foundation operated under cooperative agreement by Associated Universities, Inc.

This work is based [in part] on observations made with the Spitzer Space Telescope, which was operated by the Jet Propulsion Laboratory, California Institute of Technology under a contract with NASA. 
Herschel is an ESA space observatory with science instruments provided by European-led Principal Investigator consortia and with important participation from NASA.
Based on data products from observations made with ESO Telescopes at the La Silla Paranal Observatory under ESO programme ID 179.A-2005 and on data products produced by CALET and the Cambridge Astronomy Survey Unit on behalf of the UltraVISTA consortium.
This research has made use of the NASA/IPAC Infrared Science Archive, which is funded by the National Aeronautics and Space Administration and operated by the California Institute of Technology.

This research made use of \textsc{astropy}, a community-developed core \textsc{python} package for Astronomy \citep{astropy:2013, astropy:2018}; \textsc{matplotlib} \citep{Hunter2007_matplotlib}; \textsc{numpy} \citep{harris2020_numpy}; and \textsc{scipy} \citep{Virtanen2020_SciPy}.

\section*{Data Availability}

The ALMA data that is the focus of this article is publicly available on the ALMA science archive (2021.1.01311.S). The remaining data on COS-87259 are detailed in \citet{Endsley2022_radioAGN}.




\bibliographystyle{mnras}
\bibliography{paper_ref} 



\appendix

\section{PV Diagram}
\begin{figure}
\includegraphics{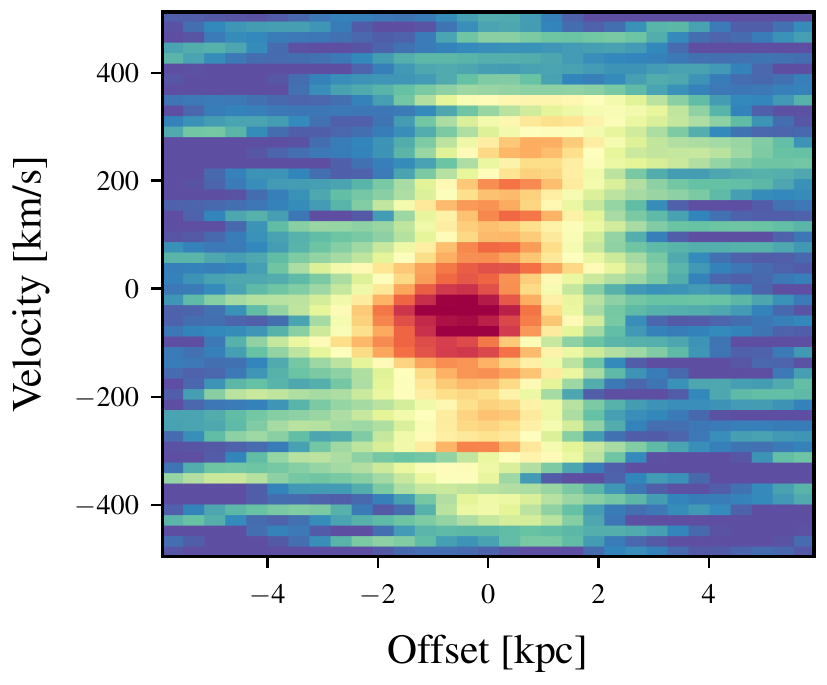}
\caption{The position-velocity diagram of [CII]158$\mu$m emission from COS-87259 extracted along the major axis. There is no clear evidence of multiple distinct components separated in velocity space.}
\label{fig:PV}
\end{figure}

We show the position-velocity diagram of [CII]158$\mu$m emission from COS-87259 in Fig. \ref{fig:PV}.

\bsp	
\label{lastpage}
\end{document}